\documentclass[reprint,aip,jcp,usenames, dvipsnames]{revtex4-1}
\draft

\usepackage[mathletters]{ucs}
\usepackage{soul}
\usepackage[english]{babel}

\usepackage{tikz,lipsum,lmodern}
\usepackage[most]{tcolorbox}

\usepackage{graphicx}
\usepackage{color}

\newcommand {\blue}[1]{{#1}}
\newcommand{\figref}[1]{Figure \ref{#1}}
\newcommand{\ie}{i.e.~}
\newcommand{\eg}{e.g.~}

\begin{document}

\title{Opportunities at the interface of network science and metabolic modelling}

\author{Varshit Dusad}
\affiliation{Department of Life Sciences, Imperial College London}
\author{Denise Thiel}
\affiliation{Department of Mathematics, Imperial College London}
\author{Mauricio Barahona}
\affiliation{Department of Mathematics, Imperial College London}
\author{Hector C. Keun}
\affiliation{Department of Surgery and Cancer, Imperial College London}
\affiliation{Department of Metabolism, Digestion and Reproduction, Imperial College London}
\author{Diego A. Oyarz\'un}
\affiliation{School of Biological Sciences,  University of Edinburgh}
\affiliation{School of Informatics, University of Edinburgh}
\affiliation{Corresponding author: d.oyarzun@ed.ac.uk}

\date{\today}

\begin{abstract}
Metabolism plays a central role in cell physiology because it provides the molecular machinery for growth. At the genome-scale, metabolism is made up of thousands of reactions interacting with one another. Untangling this complexity is key to understand how cells respond to genetic, environmental, or therapeutic perturbations. Here we discuss the roles of two complementary strategies for the analysis of genome-scale metabolic models: Flux Balance Analysis (FBA) and network science. While FBA estimates metabolic flux on the basis of an optimisation principle, network approaches reveal emergent properties of the global metabolic connectivity. We highlight how the integration of both approaches promises to deliver insights on the structure and function of metabolic systems with wide-ranging implications in discovery science, precision medicine and industrial biotechnology.
\end{abstract}

\maketitle

\section{Introduction}
Metabolism comprises the biochemical reactions that convert nutrients into biomolecules and energy to sustain cellular functions. Advances in high-throughput screening technologies have enabled the quantitative characterisation of metabolites, proteins and nucleic acids at the genome-scale, revealing previously unknown links between metabolism and many other cellular processes. For example, gene regulation\cite{Chubukov2014}, signal transduction\cite{Tretter_2016}, immunity\cite{Loftus_2016} and epigenetic modifications\cite{Reid2017} have been shown to interact closely with metabolic processes. The increasing availability of data and the fundamental roles of metabolism in various cellular phenotypes\cite{Tonn2019} have triggered a surge in metabolic research, together with a revived need for computational methods to untangle its complexity.

At the genome scale, metabolism comprises multiple interconnected reactions devoted to the production of energy and synthesis of essential biomolecules (e.g.~proteins, lipids or nucleic acids). The notion of a metabolic pathway is typically employed to organise sets of related reactions into functionally cohesive subsystems. Thus, lipid pathways, for example, are traditionally studied as distinct subsystems from amino acid or aerobic respiration pathways. Although conveniently descriptive, such \textit{a priori} partitioning  can obscure the links between other relevant layers of metabolic organization. Furthermore, metabolic connectivity is not static but actively responds and adapts to extracellular cues. Through various layers of transcriptional, translational and post-translational regulation, metabolic pathways can be activated or shut down depending on external perturbations. These metabolic shifts drive a number of fundamental biological processes, such as microbial adaptations to growth conditions~\cite{Dai_X_2016,Hartline2020} or the ability of pathogens to rewire their metabolism and evade the action of antimicrobial drugs~\cite{Olive_2016}. Metabolic adaptations are also thought to modulate the onset of complex diseases such as cancer~\cite{Hanahan_2011, Pavlova2016-jr}, diabetes, Alzheimer's, among others~\cite{DeBerardinis2012-uq,Suhre2012}. As a result, there is a growing need for computational tools that go beyond classical pathway definitions and can uncover hidden relations between metabolic components.

The complexity of metabolism has prompted the development of a myriad of methods to analyse its connectivity\cite{Wishart_2018_hmdb}. For specific pathways, kinetic models based on differential equations are widely employed to describe temporal dynamics of metabolites\cite{Steuer2006,Saa2017}. At the genome scale, however, the construction of kinetic models faces substantial challenges\cite{Srinivasan2015}. Such models require a large number of parameters, many of which have not been experimentally measured, or their values are subject to large uncertainty. As a result, the majority of genome-scale analyses are based on the metabolic stoichiometry alone. A widely adopted method for genome-scale modelling is Flux Balance Analysis\cite{Palsson2015-hp} (FBA), a powerful framework to predict metabolic fluxes on the basis of an optimisation principle applied to the network stoichiometry. Alternatively, from the stoichiometry one can build graphs, a computational description of complex systems that has become the cornerstone of network science\cite{Newman2010-rq}.

In this paper we discuss the relationship between FBA and graph-based analyses of metabolism, and we underline the complementary perspectives they bring to the understanding of metabolic organisation. On the one hand, FBA has been shown to predict metabolic activity in various environmental and genetic contexts; on the other, network science can shed light on the emergent properties of global metabolic connectivity. Both approaches share a common root in the genome-scale stoichiometry of cellular metabolism, yet they offer different tools for its analysis. In the following, we discuss their advantages and caveats, highlighting the need and opportunities for integrated methods that combine flux optimisation with network science.

\section{Flux Balance Analysis}
A large number of methods have been developed for the analysis of genome-scale metabolic networks\cite{Lewis2012}; these are generally described as \textit{constraint-based methods}~\cite{Orth2010-kg}, an umbrella term for various techniques focused on the solution of the steady state equation:
\begin{align}
    \mathbf{S}v &= 0,\label{eq:ss}
\end{align}
where $\mathbf{S}$ is the $n\times m$ stoichiometry matrix for a model with $n$ metabolites and $m$ reactions, and $v$ is a vector containing the $m$ reaction fluxes.

In general, Eq.~\eqref{eq:ss} is satisfied by an infinite number of flux vectors. A number of methods aim at probing the geometry of such flux solution space. For example, Elementary Flux Modes~\cite{Klamt_2017} and Extreme Pathways\cite{pmid16183876} are two complementary techniques for decomposing the solution space into simpler units\cite{Zanghellini_2013,Muller_2014}. Other methods for exploring the solution space include random flux sampling with Monte Carlo methods\cite{Wiback2004-MC}, the use of dimensionality reduction techniques\cite{Bhadra2018}, and various structural decompositions of the stoichiometric matrix\cite{Ghaderi2020}.

The most widespread method for genome-scale modelling is Flux Balance Analysis (FBA), which selects a vector of metabolic fluxes $v$ in Eq.~\eqref{eq:ss} as a solution to the optimisation problem:
\begin{align}
\begin{split}
	&\mathrm{max}_v\,\, J(v) \\
	\text{subject to:}\quad &\mathbf{S}v = 0\\
	&V_i^{\text{min}}\leq v_i \leq V_i^{\text{max}},\,i=1,\ldots,m,
	\end{split}\label{eq:FBA}
\end{align}
where $(V_i^{\text{min}},\, V_i^{\text{max}})$ are bounds on each flux. The objetive function $J(v)$ is chosen to describe the physiology of a particular organism under study. In microbes, biomass production is the most common choice for the objective function, in which $J(v)=c^T \cdot v$, \ie, the rate of biomass production is assumed to be a linear combination of specific biosynthetic fluxes, defined by the positive vector of weights $c$. There are many dedicated FBA software packages\cite{Lakshmanan2014,Lieven2020} and its popularity has led to a myriad of extensions\cite{Lewis2012} that account for other complexities of cell physiology such as gene regulation~\cite{Covert2008}, dynamic adaptations~\cite{Waldherr2015,Rugen2015}, and many others~\cite{Heirendt_2019}.

Flux Balance Analysis has found applications in diverse domains, including cell biology~\cite{McCloskey_2013}, metabolic engineering~\cite{Nielsen_Keasling_2016}, microbiome studies~\cite{Manor_2014, Khandelwal_2013, Rosario_2018}, and personalised medicine~\cite{Diener_2016, Nielsen_2017_personalized,Raskevicius_2018}. A salient feature of FBA is its ability to incorporate various types of `omics datasets into its predictions. Various approaches have been developed for this purpose~\cite{Yizhak2014-wp, Lee2012-pb,Colijn2009-it, Jerby2010-io, Becker2008-yt, Nam2014-lj, Agren2014-ug, Wang2012-qp}, most of which incorporate experimental data into the metabolic model through adjustments of the stoichiometric matrix $\mathbf{S}$ or the flux bounds $V_i^{\text{min}}$ and $V_i^{\text{max}}$ in~\eqref{eq:FBA}.

A popular use case of FBA is the identification of essential genes, \ie genes that severely impact cellular growth when knocked out. Through simulation of gene deletions, FBA can serve as a systematic tool for \textit{in silico} screening of lethal mutations, and identification of biomarkers and drug targets in disease\cite{Folger2011-jn,Raman2009-yh, Gatto2015-tw, Lehar2009-ou, Robinson2017-gz, Krueger2016-ol,Pagliarini2016}. A related application of FBA is the study of metabolic robustness. Since only a fraction of all metabolic reactions are essential in a given environment, knocking out non-essential reactions often has little effect on the phenotype. This is because many reactions have functional backups through other pathways, so as to preserve cellular function in face of perturbations. By providing insights into the reorganisation of fluxes under different conditions, FBA can also help improve our understanding of robustness to gene knockouts\cite{Palsson2015-hp, Larhlimi2011-nf,Blank2005-lr,Deutscher2006-zj,Ho2016-pe}, gene mutations \cite{Fong2004-di} and different growth conditions\cite{Ibarra2002-qi}.

One limitation of FBA is the crucial importance of the objective function to be optimized, which needs to be designed to represent cellular physiology. In microbes, a common choice is maximisation of growth rate, but it is questionable whether this is a realistic cellular objective across organisms or in different growth conditions~\cite{Feist2010-hf, Schuetz2007-hr, Garcia_Sanchez2014-rn}. Although the vast majority of FBA studies rely on the maximization of cellular growth, other objective functions have been proposed, including  maximization of ATP production~\cite{Nam2014-lj} and minimization of substrate uptake rate~\cite{Raman2009-yh}.

\section{Network science in metabolic modelling}
Network science represents complex systems as graphs where the nodes describe the components of the system and the edges describe interactions between components. This general description provides a framework for modelling large, interconnected systems across many disciplines, including biology, sociology, economics and  others\cite{Newman2010-rq}. Numerous works have analysed metabolism under the lens of network science. Graph-theoretic concepts such as degree distributions and centrality measures\cite{Jeong2000-io, Ma2003-qh, Wagner2001-ow} can reveal structural features of the connectivity of the overall system, while clustering algorithms can uncover substructures hidden in the network topology. Such tools can be combined with the analysis of perturbations, such as deletions of network nodes or edges\cite{Larhlimi2011-nf, Palumbo2005-aj}, which can represent changes in the environment, gene knockouts, or therapeutic drugs that target specific metabolic enzymes. Unlike FBA, in which the analysis depends on the choice of a specific objective function, network methods rely on the metabolic stoichiometry alone.

Metabolic modularity is an area where network science has shown promising results. Intuitively, a network module is a subset of the network containing nodes that are more connected among themselves than the rest of the network. Several studies have focused on the modularity of metabolic networks, and how the network modules can be used to coarse-grain the metabolic network into subunits~\cite{Ma2003-qh, Da_Silva2008-xb, Tanaka_2005, Zhao2007-ic, Kreimer2008-in}. The modules identified using network analysis have been found to capture the organization of textbook biochemical pathways while uncovering novel links and relationships between them~\cite{Ravasz2002-ld}. A recurring theme in these analyses is the bow-tie topology, whereby a metabolic network can be divided into an input component, an output component and a strongly connected internal component. This architecture aligns well with an intuitive understanding of metabolism, which  comprises nutrient uptake, waste production and secretion, and a large number of internal cycles which produce biomass and energy\cite{Ma2003-qh, Da_Silva2008-xb, Tanaka_2005, Zhao2007-ic, Kreimer2008-in, cooper2010rolebased}.

Despite its promise, however, network science has generally achieved mixed success in metabolic research. For example, from a network perspective it would be natural to expect that essential genes should be associated with high centrality scores~\cite{Jeong2001-kt,Jalili2016-xo, Raman2014-sh, Plaimas_2010}. This idea draws parallels from other domains, such as the internet and social networks, where highly central nodes are deemed critical for network connectivity. However, correlations between gene essentiality and node centrality have been so far shown to be weak, with various essential metabolites and reactions exhibiting low centrality scores\cite{Mahadevan2005-tf, Samal2006-vx}. This happens because poorly connected nodes are often the sole route for producing precursors that are essential for growth; in other words, such nodes lack a functional backup that can compensate for their loss. For example, Samal et al\cite{Samal2006-vx} showed that more than 50\% of essential reactions in \textit{Escherichia coli}, \textit{Saccharomyces cerevisiae}, and \textit{Staphylococcus aureus} are involved in such unique pathways, while other works noted that removal of poorly connected metabolites nodes can disrupt subsystems leading to failure of entire networks\cite{Mahadevan2005-tf,Winterbach2011-dm}. Other studies have attempted to resolve this problem with new network metrics specifically tailored to describe important features of metabolism\cite{Wunderlich2006-dz, Rahman2006-yh, Palumbo2005-aj, cooper2010rolebased, NaderiYeganeh2020, Kim2019}.

\begin{figure}[t]
\includegraphics{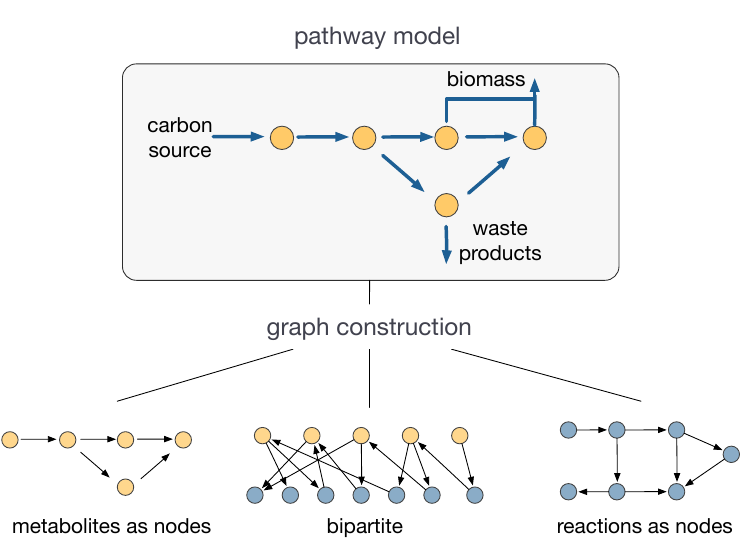}
\caption{\textbf{Graph constructions for metabolic networks.} Depending on how nodes and edges are defined, several graphs can be built from a single metabolic model\cite{Palsson2015-hp}. The conclusions drawn from graph analyses depend strongly on the choice of graph. The lack of consensus on the construction of such graphs is a key challenge for the use of network science in metabolic modelling.}\label{fig:fig1}
\end{figure}

A key challenge for the use of network science in metabolic modelling is the lack of consensus on how to build a graph from a metabolic model. For a network with $q$ nodes, the graph is encoded through the $q\times q$ adjacency matrix $\textbf{A}$, which has an entry $A_{ij}\neq 0$ if nodes $i$ and $j$ are connected, and $A_{ij}=0$ otherwise. As illustrated in \figref{fig:fig1}, depending on how nodes and edges are defined, one can build different graphs for the same metabolic model described by the stoichiometry matrix $\textbf{S}$ in~\eqref{eq:FBA}. From a metabolite-centric perspective one can build a graph where the nodes are metabolites and the edges corresponds to reactions between them\cite{Ma2003-qh, Asgari_2013}. In this case the adjacency matrix is
\begin{align}
    \textbf{A}_{n \times n} &= \hat{\textbf{S}}\hat{\textbf{S}}^T,
\end{align}
where $\hat{\textbf{S}}$ is the binary version of the stoichiometry matrix $\textbf{S}$ (i.e.~$\hat{S}_{ij}=1$ when $S_{ij}\neq0$, and $\hat{S}_{ij}=0$ otherwise). Conversely, from a reaction-centric perspective we can construct graphs with reactions as nodes and edges describing the sharing of metabolites as reactants or products\cite{Ma2004-in, Beguerisse-Diaz2018-jg}. Such graph has an adjacency matrix
\begin{align}
    \textbf{A}_{m \times m} &= \hat{\textbf{S}}^T\hat{\textbf{S}}.\label{eq:reactiongraph}
\end{align}
One can also build bipartite graphs, where both metabolites and reactions are nodes of different types~\cite{Beber_2012, Holme_2009}, or even hypergraphs where an edge connects a set of reactants to a set of products\cite{Pearcy2016-ih, Cottret_2010}. In addition, all of these graphs can be directed/undirected (when the matrix $A$ is symmetric/asymmetric), or weighted/unweighted (where the elements $A_{ij}$ can have weights encoding different properties). Such modelling choices can strongly influence the conclusions drawn from network analyses\cite{Klamt2009-mg, Bernal2011-oc, Beguerisse-Diaz2018-jg}. For example, the existence of power law degree distributions\cite{Jeong2000-io} and the small-world property in metabolism\cite{Wagner2001-ow}, two cornerstone concepts in network science, have been disputed\cite{Lima-Mendez2009-hu, Arita2004-jl} and attributed to specific ways of constructing the network graph\cite{Bernal2011-oc, Montanez2010-hy}.

A further limitation of graph-based analyses is their \textit{ad hoc} treatment of pool metabolites, e.g., H$_2$O, ATP, NADH and other enzymatic co-factors. Because pool metabolites participate in a large number of reactions, they can distort and dominate the topological properties of reaction-centric graphs\cite{Ma2003-qh}. A common approach to mitigate this problem is pruning the pool metabolites from the graph; yet there are no established best practices on how to choose which pool metabolites to prune, or how to mitigate the potential loss of information in so doing\cite{Ma2003-ah,Gerlee2009}.

Another challenge arises from the reversibility of metabolic reactions in the graph. Although all biochemical reactions are reversible, they take one direction depending on the physiological conditions. The analysis of reaction-centric graphs typically prescribe a direction for reaction flux, or they split them into forward and backward components\cite{Wagner2001-ow, Helden2002-ho}. Neither of these approaches is ideal: assigning the direction of a reaction based on one condition may not generalize across other conditions, whereas incorporating bi-directional edges increases the complexity of the analysis.

\section{Flux-weighted graphs: integration of FBA and network science}
As discussed in previous sections, both FBA and network science require modelling choices that can shape the conclusions drawn from their analyses. \blue{Tools from network science have already been employed to improve FBA pipelines in various ways\cite{Lewis2012}. Here we argue that the converse, \ie using FBA to enrich the metabolic graphs, offers promising avenues to overcome some of their individual shortcomings}. Flux information obtained from FBA solutions can be employed to assign direction and strength to the interactions between nodes in a graph. Such flux-weighted graphs allow to constrain their connectivity to various growth conditions, resulting in graphs that do not represent one universal network but are rather tailored to specific environmental or physiological contexts. As illustrated in \figref{fig:fig2}A, the integration of FBA and graph construction can thus result in a highly flexible pipeline to study metabolic connectivity in different functional states of an organism.

Although the literature on this subject is still scarce, a number of studies have demonstrated the potential of the integration of FBA into graph analyses. These studies cover a range of methodologies and applications, including \eg the identification of biomarkers\cite{Li_Jiang_2013}, detection of metabolic drug targets\cite{Li2010-ls}, and quantification of metabolite essentiality \cite{Riemer_2013,Laniau2017-qk}. In one of the early works in the subject, Smart et al\cite{Smart2008-bc} proposed an adaptation of FBA that takes into account the connectivity of individual nodes. This idea revealed new insights on how the connectivity of specific metabolites provides robustness to metabolic networks.

Other studies have explored the use of FBA to construct flux-weighted graphs with either metabolites as nodes~\cite{Yoon2007-hn, Koschutzki_2010,Riemer_2013} or reactions as nodes~\cite{Kelk_2012,Beguerisse-Diaz2018-jg, Li_Jiang_2013}. An alternative approach defined the concept of flux similarity\cite{Li2010-ls} to build reaction-drug graphs for detection of drug targets in cancer. Most recently, Hari and Lobo\cite{Hari2020} developed Fluxer, a web tool for visualization and analysis of flux-weighted metabolite graphs. The software allows the inclusion of customisable edge weights based on reaction fluxes and can perform multi-reaction knockout simulations.

In terms of applications, most studies have focused on flux-weighted graphs for the analysis of metabolic modularity and essentiality. Next we briefly discuss some of the approaches so far in these two application domains.

\subsection{Network clustering}
A promising application of flux-weighted graphs is the detection of modular subunits within genome-scale metabolic models (\figref{fig:fig2}B). The idea is that flux-weighted graphs can encode information on the strengths on interactions between graph nodes that are specific to a particular physiological state, as modelled by the FBA solution. This can potentially reveal hidden groupings within metabolism, or how known groupings change across different contexts. For example, Yoon et al\cite{Yoon2007-hn} employed experimentally determined fluxes to build flux-weighted graphs with metabolites as nodes. Using clustering algorithms on the graphs for energy metabolism of rat liver and adipose tissue formation, the approach revealed changes in cluster membership under different physiological flux distributions.

Another promising approach is the ``mass flow graph'' proposed by Beguerisse-Diaz et al\cite{Beguerisse-Diaz2018-jg}, which uses FBA solutions to weight the edges of graph with reactions as nodes. In this approach, if reaction $R_i$ produces a metabolite $x_k$ that is consumed by $R_j$, then the weight of the edge between both reactions is
\begin{align}
    w_{ij} &= \sum_k \text{(mass flow of $x_k$ from $R_i$ to $R_j$)}, \label{eq:mfg}
\end{align}
where the sum acts on all the metabolites that are produced by $R_i$ and consumed by $R_j$. The mass flows in \eqref{eq:mfg} are directly computed from the stoichiometric matrix $\mathbf{S}$ and a flux vector obtained with FBA. Different mass flow graphs can be then computed for FBA solutions corresponding to specific environmental conditions. Thanks to the flux weighting, mass flow graphs avoid the need to prune pool metabolites, a common limitation of reaction graphs\cite{Gerlee2009}. Although pool metabolites do create many connections between functionally unrelated reactions, in mass flow graphs such connections are weak as a result of the flux weighting. This feature allowed the use of multiscale community detection algorithms to study changes in the modular structure of \textit{E. coli} metabolism in various growth media\cite{Beguerisse-Diaz2018-jg}.

\subsection{Centrality and essentiality }
\begin{figure*}[t]
\includegraphics{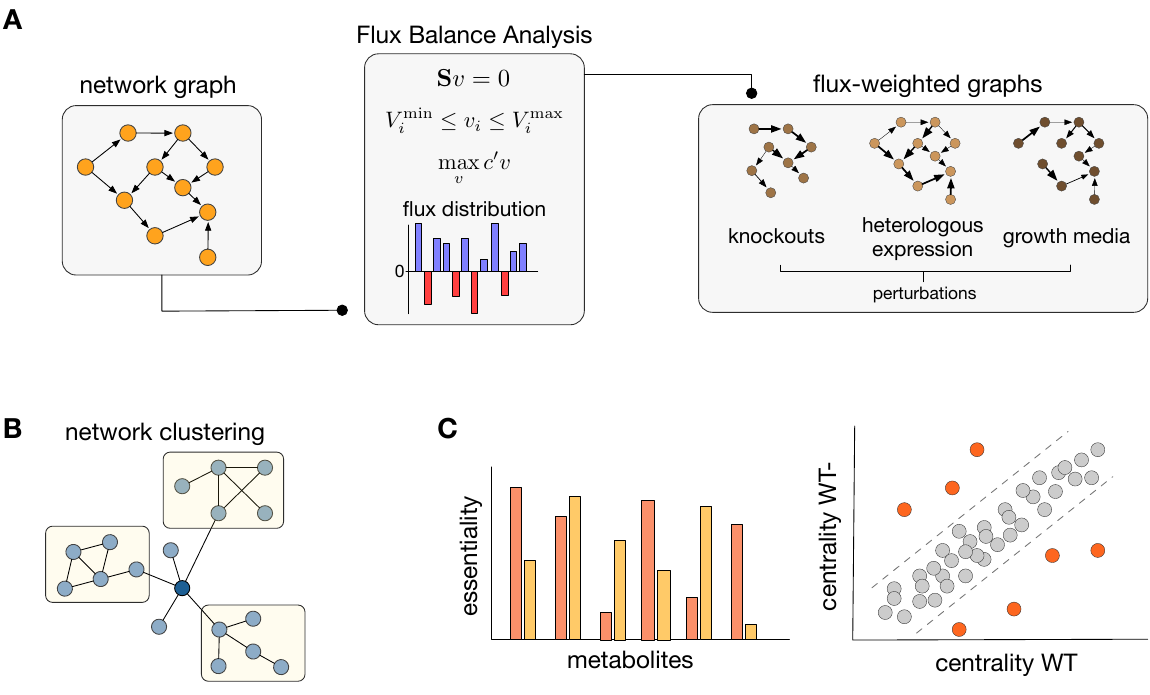}
\caption{\textbf{Integration of Flux Balance Analysis and graph theory.} (\textbf{A}) Construction of flux-weighted graphs. Starting from a network graph, FBA solutions can be used to weight the graph edges, leading to different instances of the same graph for various perturbations such as gene knockouts, heterologous gene expression, or changes in growth conditions. (\textbf{B}) Clustering of flux-weighted graphs can reveal hidden groupings in the structure of metabolic networks\cite{Yoon2007-hn,Beguerisse-Diaz2018-jg}. (\textbf{C}) Essentiality and node centrality are two areas where flux-weighted graphs offer promising potential. (Left) New essentiality scores can be defined to rank metabolites according to their impact on the phenotype\cite{Laniau2017-qk,Koschutzki_2010,Riemer_2013}. (Right) Changes in node centrality between wild type (WT) and deletion (WT-) networks can reveal the molecular players associated to disease phenotypes; orange nodes denote reactions that undergo substantial changes in centrality upon gene deletions\cite{Beguerisse-Diaz2018-jg,Li_Jiang_2013}.} \label{fig:fig2}
\end{figure*}

Flux-graph integration has also provided opportunities to explore centrality scores for quantifying essentiality of reactions and metabolites (\figref{fig:fig2}C). One example of this approach\cite{Li_Jiang_2013} demonstrated that the combination of PageRank centrality\cite{Newman2010-rq} with flux information can help to identify candidate biomarker genes in disease. The use of flux-weighted graphs also allows to compare their connectivity between models that lack specific metabolic genes, \eg in the case of mutants or genetic deficiencies found in metabolic disorders. For example, PageRank centrality was employed in conjunction with mass flow graphs\cite{Beguerisse-Diaz2018-jg} to study structural changes in hepatocyte metabolism in primary hyperoxaluria type 1, a rare metabolic disease characterised by the lack of the \textit{agt} gene involved in glyoxylate breakdown\cite{Pagliarini2016}. This approach showed that reactions which underwent the highest PageRank changes between healthy and diseased states were directly related to the PH1 phenotype (\figref{fig:fig2}C). Importantly, some of the changes in PageRank centrality did not correlate with changes in flux, providing strong evidence that metabolic graphs can encode information that cannot inferred from FBA alone.

A number of other works have sought to define new, metabolism-specific, centrality scores that can reveal new information on the topology of metabolic networks. For example Kosch\"utzki and colleagues\cite{Koschutzki_2010} built a novel ``flux centrality'' score for metabolites in networks where only the carbon exchanges are modelled as edges. This metric emphasises the role that a metabolite plays in biomass formation based on both topology and flux, penalising the impact of highly connected pool metabolites. Riemer et al\cite{Riemer_2013} combined the classic notion of metabolic branch points, \ie metabolites that are substrates to multiple downstream pathways, with reaction fluxes so as to rank metabolites according to various metrics of essentiality. A similar approach to establish metabolite essentiality was presented by Laniau et al\cite{Laniau2017-qk}, where they classify metabolites on the basis of their capacity to influence the activation of a target objective function.

\section{Discussion}
Recent discoveries have led to a renewed interest in the interplay of metabolism with other layers of the cellular machinery\cite{Chubukov2014,Tretter_2016,Loftus_2016,Reid2017,Tonn2021}. Due to the complexity and scale of metabolic reaction networks, computational methods are essential to tease apart the influence of metabolic architectures on cellular function. Here we have discussed the complementary roles of Flux Balance analysis and network science in the analysis of metabolism at the genome scale. Although both approaches start from the metabolic stoichiometry, they differ in their mathematical foundations and the type of predictions they produce. FBA predictions can be accurate but their effectiveness requires high quality `omics datasets. Network science, in contrast, requires nothing more than the metabolic stoichiometry, yet can lead to misleading predictions depending on how the network graph is built. As a result, so far FBA has led to more successful connections with experimental results than network science.

When used in isolation, both FBA and network science can be insufficient to understand changes in metabolic connectivity triggered by physiological or environmental perturbations. Here we argue that the use of flux-weighted graphs (\figref{fig:fig2}A) allows for a natural integration of FBA and network science, applicable in many subject domains. For example, with the rise of big data in the life sciences, there is a growing interest in using patient metabolic signatures to tailor treatments\cite{ODay2018}. Computational methods can play a key role in detecting drug targets involved in metabolic activity, and how their targeting can disrupt metabolic connectivity. A particularly promising area is cancer treatment, where there is considerable interest on drugs that target specific metabolic enzymes\cite{Neradil_2012, Nishi_2016}. Moreover, novel data-driven approaches based on machine learning can also be integrated with FBA\cite{pmid32444610,Zampieri2019} and network science to extend their capabilities into novel applications.

Another promising application domain is industrial biotechnology\cite{DeLorenzo2018}, where so called ``microbial cell factories'' are engineered for production of commodity chemicals and fine products\cite{Lee2012}. In this field, FBA is widely employed for strain design, with the goal of finding combinations of genetic interventions that maximise production of a desired metabolite. A recent trend is to increase production with synthetic biology tools and dynamic control of gene expression~\cite{Brockman_2015,Liu2018}. This approach needs computational methods that capture the dynamic reallocation of metabolic flux. Integrating FBA solutions with network models can provide a versatile tool to identify suitable genetic modifications for microbial strains with increased production.

Further developments at the interface of FBA and network science offer a novel way to explore the impact of perturbations on metabolic connectivity. The flexibility of FBA allows for the modelling of metabolic perturbations of various kinds, including changes in growth conditions, deletion of metabolic genes or the action of enzyme inhibitors, whereas the application of network-theoretical tools can bring a broadened understanding of emergent properties of the overall system. This flexibility offers promising potential to deploy network science tools across a range of questions in basic science, biomedicine and industrial biotechnology.

\section*{Acknowledgments}
This work was supported by Cancer Research UK (C24523/A27435), the Cancer Research UK Imperial Centre, and the EPSRC Centre for Mathematics of Precision Healthcare (EP/N014529/1).

\bibliography{references.bib}

\end{document}